\documentclass[12pt,titlepage]{article}

\usepackage{amsmath,amssymb,amstext}
\usepackage[pdftex]{graphicx}
\usepackage{xspace,colortbl}
\usepackage{multirow}

\begin{document}

\normalsize

\newcommand{\dt}[1]{\mbox{\rm d}#1}
\newcommand{\Rd}[2][R]{\mathcal{#1}^{#2}}
\newcommand{\Cal}{\mathcal{X}}
\newcommand{\beq}{\begin{equation}}
\newcommand{\eeq}{\end{equation}}
\newcommand{\e}{\mbox{e}}

\title{\bf {\Large   Scalable Bayesian Multiple Changepoint Detection via Auxiliary Uniformization \\ }}

\vspace{3cm}
\author{\sc\large { Lu Shaochuan} \\ School of Statistics, Beijing Normal University,\\ Xin Jie Kou Wai Da Jie 19, \\Beijing, P. R. China
\\lvshaochuan@bnu.edu.cn\\tel:+86-10-82075411
}

\maketitle

\begin{abstract}
    By attaching auxiliary event times to the chronologically ordered observations, we formulate the Bayesian multiple changepoint problem of discrete-time observations into that of continuous-time ones. A version of forward-filtering backward-sampling (FFBS) algorithm is proposed for the simulation of changepoints within a collapsed Gibbs sampling scheme. Ideally, both the computational cost and memory cost of the FFBS algorithm can be quadratically scaled down to the number of changepoints, instead of the number of observations, which is otherwise prohibitive for a long sequence of observations. The new formulation allows the number of changepoints accrue unboundedly upon the arrivals of new data. Also, a time-varying changepoint recurrence rate across different segments is assumed to characterize diverse scales of run lengths of changepoints. We then suggest a continuous-time Viterbi algorithm for obtaining the Maximum A Posteriori (MAP) estimates of changepoints. We demonstrate the methods through simulation studies and real data analysis .\\

{\sl keywords: Multiple changepoint problems; Poisson randomization; Forward-filtering backward-sampling algorithms; Collapsed Gibbs sampler; Infinite hidden Markov models; Viterbi algorithms}
\end{abstract}
\cleardoublepage

\section{Introduction}
 Chronologically ordered data streams are often heterogeneous rather than homogeneous. Multiple changepoint models are used in this scenario for splitting the data streams into a (random) number of subsets, so that observations within the same grouping are regarded as homogeneous and observations across different groupings arise from different data generating mechanisms. These models are widely applied in signal processing, DNA segmentation in bioinformatics, climate analysis, geophysical research and change of volatilities in finance market, among many other important applications. For these models, the primary interest lies in making inference on the number of changepoints and their locations, including the evaluation of uncertainties about them.

 The problem is pioneered by Page (1954). Since then, many approaches have been developed in this area and it continues to thrive, prompted by both statisticians and many other communities in important applications. A full list of different methods in real application is beyond our reach. Currently, PELT (Killick et al. 2012), Wild Binary Segmentation (Fryzlewicz 2014) and SMUCE (Frick et al. 2014) are widely applied. Generally, changepoints are detected either in a retrospective approach or in a realtime manner. In Bayesian retrospective changepoint detection, much research is based on the use of Monte Carlo strategies, see Lavielle and Lebarbier (2001) . For the inference of a fixed number of changepoints, Stephens (1994) and Chib (1998) suggest two versions of Bayes hierarchical models for changepoints, and two Gibbs sampling schemes are suggested correspondingly. For multiple changepoint models with unknown number of changepoints, reversible jump MCMC approach is introduced in Green (1995). Within a framework of product partition models, Fearnhead (2006) suggests an efficient Monte Carlo strategy via forward-backward recursions to simulate the random changepoints, see also Barry and Hartigan (1992, 1993) and Ruggieri and Lawrence (2014). In an alternative approach, Giordani and Kohn (2008) build a normal state-space representation for the breaking process with random number of structural breaks via a "mixture innovation models" . Ko et al. (2015) introduce a Bayesian nonparametric multiple changepoint model via Dirichlet process priors, in which the number of changepoints can accrue unboundedly. Peluso et al. (2019) develop another Bayesian nonparametric (semiparametric) approach to modelling the changepoint process, in which a Dirichlet process mixture prior is elicited for the observations and the changepoint recurrence mechanism follows a Markov process with time-varying transition rate matrices. Online detection of changepoints via particle filtering is developed in Chopin (2007), Fearnhead and Liu (2007) and Yildirim et al. (2013), among some others.

 This paper propose a new model for Bayesian retrospective multiple changepoint detection in discrete time. The proposed multiple changepoint model is a hidden Markov model on countable infinite state space with a left-to-right transition probability matrix. The multiple changepoint model is flexible enough to allow the number of changepoints accrue unboundedly upon the arrivals of new data, in contrast to the multiple changepoint model with the number of changepoints fixed or upper-bounded. Often, the number of changepoints is evaluated by fitting a range of multiple changepoint models with diverse number of changepoints fixed. After that, a model selection criterion is applied to choose the number of changepoints, see Davis et al. (2006) and Zhang and Siegmund (2007). However, the definition of model complexities in different changepoint  selection criterion can be sharply different. It is not straightforward to evaluate the uncertainties of the number of changepoints and their locations in this case. The proposed new model formulation also allows the changepoint recurrence mechanism varies from segment to segment, avoiding potential model bias introduced by assuming a constant changepoint recurrence rate.

 We propose an efficient Gibbs sampling scheme at low computational cost and memory cost. The Gibbs sampling scheme is based on a new version of forward-filtering backward-sampling (FFBS) algorithm to sample approximately from the posterior of the number of changepoints and their positions. For multiple changepoint models with a random number of changepoints, often, the changepoints are detected in an event-by-event scale, see e.g. Fearnhead (2006), Ruggieri and Lawrence (2014) and Peluso et al. (2019). In this case, both the computational cost and memory cost are quadratic to the number of observations, which is prohibitive for large $N$. Often, pruning skills or particle filtering methods are applied to deal with the problem of inflated computational cost and memory cost, see Fearnhead (2006) and Fearnhead and Liu (2007). We suggest a FFBS algorithm for the simulation of changepoints directly at the scale of the number of changepoints. The computational cost and the memory cost of the new FFBS algorithm can be scaled down quadratically to the number of changepoints, rather than the number of observations, which is ideal for the detection of changepoints in a long time series with sparsely distributed changepoints. The new FFBS algorithm is based on a randomized blocking technique for the discrete-time observations and the locations of changepoints are detected in a batch-by-batch manner. In contrast to the approach of blocking observations in fixed size, which will bring in blocking errors, there is no blocking errors to be accounted for in this case. The randomized blocking technique is facilitated by introducing an auxiliary event times in continuous time to the chronologically ordered discrete observations and utilizing the uniformization scheme to re-discretization (regrouping) the observations. Recently, the uniformization technique is effectively applied in Bayesian inference for Markov modulated Poisson processes and continuous-time Bayes network (Rao and Teh, 2013).

 In section 2, we formulate the multiple changepoint model of the discrete-time observations into a continuous-time infinite hidden Markov model. We then introduce a continuous-time FFBS algorithm for the simulation of the number of changepoints and their locations within a Gibbs sampling scheme in section 3. Although the set of changepoints can be summarized from the posterior samples of the latent Markov chain, we argue that it is desirable to obtain the maximum a posteriori (MAP) estimates of the changepoints in this scenario. In section 4, we discuss a continuous-time version of Viterbi algorithm for the retrieval of the most likely trajectory of the latent Markov chain, i.e. the set of changepoints. We demonstrate the methods by 4 numerical examples in section 5. The first numerical example displays the scaling effects of the tuning parameter in Algorithm 3 on the accuracies and efficiencies of the estimation for changepoints. The second numerical simulation demonstrates the multiple changepoint detection for a relatively long sequence of exponentially distributed observations with sparsely distributed changepoints. The third numerical example is a real data analysis for the time-varying patterns of New Zealand deep earthquakes. For this numerical example, large uncertainties appear for the number of changepoints and their locations. The last numerical example is the detection of changepoints for the well-log data set, which is also analysed in many other studies.

\section{Model Formulations and Notations}
 Let $\mathrm{y}_{1 : n}=\left(y_{1}, y_{2}, \dots, y_{n}\right)$ be a sequence of discrete-time observations from a distribution $\mathbb{F}_{\theta}(y)$. We assume the observations are subject to abrupt changes at unknown locations $1\triangleq\tau_0<\tau_1<\cdots<\tau_m<\tau_{m+1}\triangleq n$, such that $\mathrm{y}_{1 : n}$ is partitioned into $m+1$ segments by $m$ changepoints $\mathbf{\tau}=(\tau_1, \cdots, \tau_m)$. The number of changepoints is also unknown. The main interest lies in making inference for the number of changepoints, their locations and the model parameters $\mathbf{\theta}=(\theta_1, \cdots, \theta_{m+1})$ of $\mathbb{F}_{\theta}(y)$ in each segment. In Bayesian context, priors for the number of changepoints, their locations and the model parameters in each segment need to be specified. One approach to the prior elicitation is to define hierarchically a prior for the number of changepoints $m$ and a conditional prior for their locations conditioned on $m$, see e.g. Green (1995). Another approach to prior elicitation is to implicitly define the priors for the number of changepoints and their locations by specifying the run lengths between adjacent changepoints, see Chopin (2007), Fearnhead (2006) and Fearnhead and Liu (2007). In this paper, we choose the later approach for reasons illuminated in the following sections. For Bayes multiple changepoint detection, it is popular to model the changepoint recurrence as a hidden Markov process and the posterior inference is facilitated by forward-backward recursions and Viterbi algorithm. In Fearnhead (2006), Fearnhead and Liu (2007) and Ruggieri and Lawrence (2014), it is noted that both the memory cost and computational cost of the forward-backward recursions or the dynamic programming algorithm are quadratic to the number of observations, which is prohibitive when the data set is large and the changepoints are only sparsely distributed. Often, particle filters or pruning skills have been applied to deal with these problems.

  This paper discuss Bayesian multiple changepoint detection within the framework of hidden markov models. To deal with the issue of computational cost and memory cost of FFBS algorithm, one probably should avoid to compute and store the filtering probabilities of the changepoint process on an event-by-event scale. One may consider to block the observations, then compute and store the filtering probabilities of the changepoint process on a block-by-block scale. However, it is lack of principled approach to determine the block size. Assuming a constant blocking size is certainly insufficient to accurately locate changepoints since the changepoint recurrence rate may be highly variable. We consider a randomized blocking strategy by treating the sequence of discrete-time observations in a continuous time framework. Then the ordered observations, attached with event times, are re-discretized via uniformization strategies. Through this randomized blocking strategy, the changepoints may be detected on a block-by-block scale via discrete-time FFBS algorithm.  Assume $\mathrm{y}_{1 : n}$ is observed at $(t_1, \cdots, t_n)$ over an artificial time interval $[0,T]$. Note that there is no any loss of information on the number of changepoints and their locations by introducing an artificial continuous-time interval and placing the discrete-time observations on this interval sequentially in an arbitrary way, so that the sequence of observations $\mathrm{y}_{1 : n}$ are attached with auxiliary observational times. The benefit of formulating the Bayesian multiple changepoint problems in a continuous-time framework is mainly computational. By a randomized time-discretization via uniformization, we can deal with the Bayesian multiple changepoint detection directly at the scale of changepoint numbers, leading to lowered computational cost and memory cost. Unlike the usual time-discretization for a continuous-time random process, there is no discretization error need to be accounted for. Let $m$ changepoints are located at $\tau_1<\cdots<\tau_m$ in $[0,T]$. The exact locations of changepoints can be backtracked into serial numbers between 1 and n.

  We assume the run length of the $i$-th segment is exponentially distributed with a rate parameter $q_i$. This type of priors for the recurrence of  changepoints is a latent Markov chain $\mathbf{X}(t)$ defined on a numerable infinite state space, with a left-to-right transition rate matrix $\mathbf{Q}$ given by
 \[\mathbf{Q}=
\begin{pmatrix}
-q_1 & q_1 & 0 & \cdots & 0 & 0 & 0& \cdots\\
0 & -q_2 & q_2 & \cdots & 0 & 0 & 0& \cdots\\
\cdots & \cdots & \cdots & \cdots & \cdots &\cdots &\cdots\\
0 & 0 & 0 & \cdots & -q_m & q_m & 0 & \cdots\\
\cdots & \cdots & \cdots & \cdots & \cdots &\cdots &\cdots
\end{pmatrix}.
\]
 Upon $\mathbf{X}(t)$ sojourns in the $i$-th state, the observations are distributed according to $\mathbb{F}_{\theta_i}(y)$. This type of formulation for the multiple changepoint process is flexible in that it allows varying scale of run lengths of changepoints to be characterized in scenarios of a highly variable segment lengths between changepoints, so that potential model bias by assuming a constant changepoint recurrence rate is avoided. Current model formulation also allows the number of changepoints accrues unboundedly with the arrivals of new data, which is flexible in contrast to the multiple changepoint model with the number of changepoints fixed or upper bounded.

  In the following, the bracket $[j,i]$  after a matrix $A$ denotes the $(j,i)$-th entry of $A$. All the observations of $\mathbf{Z}(t)$ over a time interval $[a,b]$ or $[a,b)$ is denoted by $\mathbf{Z}[a,b]$ or $\mathbf{Z}[a,b)$ respectively. Generically, $p(.)$ denotes a probability density (mass) function. For convenience,  $\mathbf{Y}_{\theta}(t)$ is abbreviated to $\mathbf{Y}(t)$. Denote the inter-event time $t_i-t_{i-1}$ by $\Delta t_i$. Throughout the discussion, we may denote a vector $(z_1, \cdots, z_n)$ by $z_{1:n}$.

\section{Scalable Bayesian Multiple Changepoint Detection}
\subsection{Uniformization}
  Let $\mathbf{Q}=(q_{ij})$ be a transition rate matrix of a continuous-time, irreducible Markov chain $\mathbf{X}(t)$ on a countable infinite state space $\mathbb{S}$ with transition rates $q_{i}=\sum\limits_{j\neq i}q_{ij}.$ Assume the transition rates are uniformly upper bounded by some constant $\lambda<\infty$, such that
  \begin{equation}
   q_{i}=\sum\limits_{j\neq i}q_{ij} < \lambda.
  \end{equation}
  Define a uniformized transition matrix $\mathbb{P}$ by
  \begin{eqnarray}
  \mathbb{P}[i,j]=
  \begin{cases}
  \frac{q_{ij}}{\lambda}, \quad\quad\quad\quad if\, j\neq i; \\
  1-\sum\limits_{k\neq i}\frac{q_{ik}}{\lambda}, \quad if\, j=i, \\
  \end{cases}
  \end{eqnarray}
  which can be written by $\mathbb{P}=I+\mathbf{Q}/\lambda$. It is shown that, for all $i,j\in \mathbb{S}$ and $t>0$,
  \begin{equation}
    \mathbb{P}_t[i,j]=\sum_{k=0}^{\infty}\frac{(t\lambda)^k}{k!}e^{-t\lambda}\mathbb{P}^k[i,j],
  \end{equation}
  where $\mathbb{P}^k$ is the $k$-th matrix power of the transition probability matrix $\mathbb{P}$. The idea of "Uniformization" is first introduced by Jensen (1953), see Van Dijk et al. (2018) for a recent overview of uniformization methods. The equation has a natural interpretation as Poisson randomization. The jumps of $\mathbf{X}(t)$ occur in terms of a Poisson process with rate $\lambda$. Given $k$ Poisson event times, the jump times are uniformly distributed over the time interval. Among the jump times, the state transition may not happen, which corresponds to a dummy transition or a virtual transition from  $i \, to \, i$. The state transitions happen according to a discrete-time Markov chain with the transition probability matrix given by (2). The uniformization scheme is also applicable for continuous-time Markov chain $\mathbf{X}(t)$ on a numerable infinite state space or time-inhomogeneous Markov chain, see Van Dijk et al.(2018).

\subsection{Continuous-time FFBS Algorithm}

 Assume the observations $\mathrm{y}_{1 : n}$ are partitioned into $m+1$ segments by $m$ changepoints $\tau_1,\tau_2,\cdots,\tau_m$. Within the $i$-th segment, observations from $\mathbf{Y}(t)$ is distributed according to $\mathbb{F}_{\theta_i}(y)$. The likelihood is thus written by
   \begin{equation*}
   \mathbb{L(}m,\mathbf{\tau},\mathbf{\theta})=p(\mathbf{Y}[0,T]|m,\mathbf{\tau},\mathbf{\theta})=\prod_{i=1}^{m+1}p(\mathbf{Y}[\tau_{i-1}, \tau_i)|\theta_i).
   \end{equation*}
 In following discussions, we treat the model parameters $\theta_i$ as nuisance parameter and integrate it out from the likelihood. After specifying priors for $\theta$, the marginal likelihood of $\mathbf{Y}(t)$ on an interval $[s,t)$ is given by
  \begin{equation*}
   \mathcal{M}(s,t)=\int_{\Theta} p(\mathbf{Y}[s,t)|\theta)p(\theta)\,d\theta.
  \end{equation*}
  The marginal likelihood can be evaluated exactly upon the prior of $\theta$ is chosen from conjugate ones or some numerical integration procedure is applied. In this scenario, we suggest a collapsed Gibbs sampling scheme to improve the efficiencies of Monte Carlo sampling. In current model formulation, the prior of m changepoints located at $\tau_1, \cdots, \tau_m$ is given by $\prod_{i=1}^{m} q_ie^{-q_i(\tau_i-\tau_{i-1})}e^{-q_{m+1}(\tau_{m+1}-\tau_{m})}$. The marginal posterior of the number of changepoints and their locations is written by
   \begin{equation}
   p(m,\tau|\mathbf{Y}[0,T])\propto \prod_{i=1}^{m}\{q_ie^{-q_i(\tau_i-\tau_{i-1})}\mathcal{M}(\tau_{i-1}, \tau_i)\}e^{-q_{m+1}(\tau_{m+1}-\tau_{m})}\mathcal{M}(\tau_{m}, \tau_{m+1}).
  \end{equation}

  Let $0\triangleq u_0<u_1 < u_2 < \cdots < u_K< u_{K+1}\triangleq T$ be a sequence of uniform times of the latent Markov chain $\mathbf{X}(t)$, such that $\mathbf{X}(u_1), \cdots, \mathbf{X}(u_t)$ forms a discrete-time Markov chain with the transition probability matrix
  \begin{equation}
  \mathbb{P}=
\begin{pmatrix}
  p_1 & 1-p_1 & 0 & \cdots & 0 & 0 & \cdots\\
0 & p_2 & 1-p_2 & \cdots & 0 & 0 & \cdots\\
\cdots & \cdots & \cdots & \cdots & \cdots & \cdots & \cdots\\
0 & 0 & 0 & \cdots & p_m & 1-p_m & \cdots\\
\cdots & \cdots & \cdots & \cdots & \cdots  & \cdots & \cdots
\end{pmatrix}.
  \end{equation}
  The number of changepoints and their locations are exactly chosen from the set of uniform times $\{u_1, u_2,  \cdots,  u_K\}$. Note that $K$ uniform times split $[0,T]$ into $K+1$ subintervals. We treat all the observations within a subinterval as a single block. Denote $\mathbf{Y}_i\triangleq \mathbf{Y}[u_{i-1},u_i)$ as the $i$-th block including all $y_j$ within $(u_{i-1}, u_i]$. Let $s_i\triangleq\mathbf{X}(u),\, u\in(u_{i-1}, u_i]$ for $i=0,1,...,K+1$ and let $g(i,k)$ denotes the probability mass function of the run length of a segment, starting from $u_i$ and terminating at $u_k$, so that
  \begin{equation}
   g(i,k)=(\prod\limits_{j=i}^{k-1} p_{s_j})(1-p_{s_k}), \quad\quad k>i.
  \end{equation}
  The associated cumulative distribution function of the run length of this segment, which is starting from $u_i$,  is written by
  \begin{equation*}
  \mathbb{G}(i,k)=\sum_{j=i+1}^{k}g(i,j), \quad\quad k>i.
  \end{equation*}
  Let $C_t$ is a state variable taking values in $0, 1,2,\cdots, t-1$, which is defined to be the location of the most recent changepoint before time $t$. If there is no changepoint before $t$, $C_t=0$. $\{C_t\}$ is a discrete-time Markov chain on $0, 1, 2,\cdots, t-1$. In Fearnhead and Liu (2007), the transition probability of $C_t$ is given by
  \begin{eqnarray}
  \mathbb{P}(C_{t+1}=j|C_t=i)=
  \begin{cases}
  \frac{1-\mathbb{G}(i, t+1)}{1-\mathbb{G}(i,t)}, \quad\quad if\, j=i; \\
  \frac{g(i,t+1)}{1-\mathbb{G}(i,t)}, \quad\quad\quad if\, j=t; \\
  0, \quad\quad\quad\quad otherwise.
  \end{cases}
  \end{eqnarray}

    The forward filtering algorithm gives the filtering probabilities
  \begin{equation}
  \mathbb{ P}(C_{t+1}=i|\mathbf{Y}_{1:t+1}) \propto
   \begin{cases}
   \mathbb{P}(C_{t}=i|\mathbf{Y}_{1:t})\left(\frac{1-\mathbb{G}(i,t+1)}{1-\mathbb{G}(i,t)}\right)\left(\frac{\mathcal{M}(i, t+1)}{\mathcal{M}(i, t)}\right), \quad\quad for\, i=0,1, \cdots, t-1;\\
   \mathcal{M}(t,t+1)\sum\limits_{j=0}^{t-1}\mathbb{P}(C_{t}=j|\mathbf{Y}_{1:t})\left(\frac{g(j,t+1)}{1-\mathbb{G}(j,t)}\right), \quad\quad for\, i=t,
   \end{cases}
  \end{equation}
 where $\mathcal{M}(i, j)$ is the marginal likelihood of $\mathbf{Y}(t)$ on $(u_{i-1}, u_j]$ .

   Upon the filtering probabilities are stored, the backward sampling step samples from the posteriors of the number of changepoints and their locations. The last changepoint $C_{K+1}$ is simulated according to the filtering probabilities $\mathbb{P}(C_{K+1}|\mathbf{Y}[0,T])$. Given the latest changepoint $C_{K+1}=t$, the next changepoint is sampled recursively and backwardly according to

  \begin{equation}
    \mathbb{P}(C_t=i|\mathbf{Y}_{1:K+1}, C_{K+1}=t)\propto \mathbb{P}(C_t=i|\mathbf{Y}_{1:t}) \left(\frac{g(i, t+1)}{1-\mathbb{G}(i,t)}\right).
  \end{equation}
  The recursion terminates when $C_t=0$.

  This completes the new version of forward-filtering backward-sampling algorithm, proposed particularly for the multiple changepoint problems in discrete time. The FFBS algorithm bears some appealing properties. Firstly, both the computational cost and the memory cost of this version of FFBS algorithm are only quadratic to the number of uniform times $K$, instead of the number of observations, which is otherwise prohibitive when the filtering probabilities are computed and stored in an event-by-event scale. The current method computes the filtering probabilities and store them directly at the scale of changepoint numbers. It is potentially suitable for dealing with multiple changepoint problems of a long sequence of observations. Secondly, when the transition intensity rate matrix $\mathbf{Q}$ of $\mathbf{X}(t)$ is known, the changepoints may be simulated directly by an exact Monte Carlo sampling via FFBS algorithm. Otherwise, the FFBS algorithm can be incorporated into a two-block Gibbs sampling scheme to sample approximately from the posterior of the number of changepoints and their locations. The two-block Gibbs sampling scheme completes by sampling the transition intensity rate matrix $\mathbf{Q}$ conditioned on the full path of the latent Markov chain.

  We assume a conjugate prior $\Gamma(a,b)$ for $q_i$. The joint prior of $\mathbf{Q}$ is given by
  $\prod\limits_{i=1}^m\frac{b^a}{\Gamma(a)}q_i^{a-1}\exp\{-bq_i\}.$
  The hyperparameters $a$ and $b$ in the priors are often specified manually. The likelihood of $\mathbf{X}(t), 0\le t\le T$ is written by
  $ \prod\limits_{i=1}^mq_i\exp\{-q_i(\tau_i-\tau_{i-1})\}. $
  Therefore, the posterior distribution of $\mathbf{Q}$ is
  \begin{equation}
  p(\mathbf{Q}|\mathbf{X}[0,T])\propto \prod\limits_{i=1}^m\frac{b^a}{\Gamma(a)}q_i^{(a+1)-1}\exp\{-[b+(\tau_i-\tau_{i-1})]q_i\}.
  \end{equation}
  From the above equation, the full conditional of $q_i$ is obviously $\Gamma(a+1,b+(\tau_i-\tau_{i-1}))$. This completes a two-block Gibbs sampler.

  \textbf{Algorithm 1} Block Gibbs sampler\\
  1. Attach auxiliary event times to the observations and turn the multiple changepoint problems in discrete time into that of continuous-time ones;\\
  2. Specify an initial value for the number of changepoints, their locations and the corresponding transition intensity rate matrix;\\
  3. Simulate a sequence of Poisson events $u_1, u_2, \cdots, u_K$ from a stationary Poisson process with a constant intensity rate $\lambda$ satisfying (1). Define a discrete-time Markov chain with the transition probability matrix $\mathbb{P}$ according to (2) on $u_1, u_2, \cdots, u_K$;\\
  4. Sample a trajectory of the latent Markov chain $\mathbf{X}(t)$ via the FFBS algorithm;\\
  5. Conditioned on the number of changepoints and their locaitons, i.e. the trajectory of $\mathbf{X}(t)$, sample the transition intensity rate matrix $\mathbf{Q}$;\\
  6. Repeat 3-5 until the final iteration reached. Backtrack the serial numbers of changepoint locations.\\

  Note that both the memory cost and the computational cost of this two-block Gibbs sampler are quadratic to the number of uniform times, which is determined by the Poisson rate $\lambda$ in the uniformization scheme. The algorithm is suitable only when the run lengths between consecutive changepoints are nearly regular. When the "changepoint recurrence rate" $q_i$s in  $\mathbf{Q}$ are highly variable, the Poisson rate $\lambda$ in the uniformization scheme will be dominated by the largest one, leading to the generation of excessive number of uniform times. The computational cost and memory cost will be sharply inflated. In this scenario, we consider to replace the constant intensity rate in (1) by a nonhomogeneous Poisson intensity rate $\lambda(t)$. Obviously, a straightforward choice is the nonhomogeneous Poisson process with piecewise constant intensity function, as suggested in Rao and Teh (2013) for Bayesian inference of the Markov modulated Poisson process and continuous-time Bayesian network. After squeezing down the Poisson intensity rate piecewisely, the number of uniform times in the uniformization scheme can be reduced, leading to a FFBS algorithm with lower memory cost and computational cost. To simulate a nonhomogeneous Poisson process, it is standard to either simulate a Poisson process piecewisely or apply the thinning method (Lewis and Shedler 1979). The thinning algorithm for a general point process with bounded conditional intensity $\lambda(t)$ is given as follows:

  \textbf{Algorithm 2} Lewis-Shedler Thinning Algorithm\\
  1. Let $U=\varnothing$. Simulate $v_1, v_2, \cdots, v_m$ according to a Poisson process with rate $\lambda$ satisfying $\lambda(t) < \lambda$ on $[0,T]$. For example, we can simulate successive interval lengths from i.i.d. exponential variables with mean $\frac 1{\lambda}$;\\
  2. For $i=1, \cdots, n$, evaluate $\lambda(v_i)$ and simulate a $U_i$ from $\mathcal{U}[0,1]$, if $U_i< \frac{\lambda(v_i)}{\lambda}$, let $U=U\cup t_i$;\\
  3. Output the remaining points $\{u_k\}_{k=1}^K$.\\

  The step 3 in Algorithm 1 is modified as follows:\\
  3'. Simulating a sequence of Poisson events $u_1, u_2, \cdots, u_K$ from a nonhomogeneous Poisson process with a piecewise constant intensity rate via Algorithm 2. Compute the transition probabilities according to (2).\\
  In simulation studies, it is observed that consecutive uniform times with small spacings are often simulated as individual changepoints, which nevertheless should be treated as a single changepoint. A scrutiny reveals that it happens when the filtering probabilities of the locations of the most recent changepoints are not concentrated on single uniform times, but rather spreading across several consecutive uniform times. It means that relatively large uncertainties exist for the locations of most recent changepoints. In this case, the backward-sampling step would treat part of them, sometimes even all of them as independent changepoints. We call it as a "knot" since all of them should be regarded as a single changepoint.  Without pruning these knots, it is obvious that the number of uniform times generated from the uniformization scheme in the next iteration will increase, causing problems such as sharply inflated memory costs and computational costs, reducing the efficiencies of the algorithm. A straightforward approach to pruning these knots is to repeat the backward-sampling step several times and choose the furthest one as the most recent changepoint before $t$. The step 4 in Algorithm 1 is modified as follows:\\
  4'. Given the simulated uniform times $\{u_k\}_{k=1}^K$, the associated state transition probabilities $\{p_{s_k}\}_{k=1}^K$ and the current values of other model parameters, calculate the forward probabilities recursively according to (8). In backward-sampling step, sample the most recent changepoint location before $t$ repeatedly and choose the furthest one as the most recent changepoint. The backward-sampling step stops when $t=0$. \\

  This straightforward approach to pruning knots among changepoints will possibly lead to slightly biased estimation of the locations of changepoints, see simulation studies in the following sections. However, this is a price paid for otherwise sharply inflated memory costs and computational costs. In real applications, the model parameters $\theta$ might be also of interest, besides for the number of changepoints and their locations. We simulate the posterior of $\theta$ by a collapsed Gibbs sampler, in which $\theta$ is sampled according to their full conditionals after the convergence of the Gibbs sampler mets, see Liu (1994).

  In step 3 of Algorithm 1, as a countermeasure to the complete randomness of the uniform times generated from uniformization scheme, it is better to keep the set of changepoints simulated from the last FFBS step into the next iteration. By keeping the set of changepoints simulated from the previous FFBS step into the next uniformization step, the most probable locations of changepoints are likely to "survive" in the following iterations, which may improve the accuracies of the estimation for the locations of changepoints. We summarize all major modifications to Algorithm 1 into the following Algorithm 3:

 \textbf{Algorithm 3} Collapsed Gibbs sampler\\
  1. 1. is same as Algorithm 1.\\
  2. Specify initial values $m, \{\tau_i\}_{i=1}^m, \mathbf{Q}$ for the number of changepoints, their locations and the corresponding transition intensity rate matrix respectively. The Poisson intensity rate $\lambda(t)$ in the uniformization scheme is assumed $k$ times of $q_i$ on $[\tau_{i-1}, \tau_i)$, $i=1,\cdots,m+1.$ \\
  3. By use of thinning methods (Algorithm 2), simulate a sequence of uniform times $\mathcal{U}'=\{u_i\}_{i=1}^{K'}$ from a Poisson process with a piecewise constant intensity rate $\lambda(t)$, which is specified $k$ times of $q_{X(t)}$ as aforementioned. Adding $\tau$ into $\mathcal{U}'$ to form a sequence of uniform times $\mathcal{U}=\{u_i\}_{i=1}^K$. Calculate the state transition probabilities $p_{s_i}, i=1,\cdots,K$ according to (2).\\
  4. Calculate and store the filtering probabilities recursively according to (8). In the backward-sampling step, simulate the most recent changepoint locations repeatedly and choose the furthest one as the most recent changepoint before $t$ according to (9). Stops at $t=0$ and obtain a new set of changepoints $\{\tau_i\}_{i=1}^m$.\\
  5. Conditioned on the number of changepoints and their locaitons, i.e. the trajectory of $\mathbf{X}(t)$, sample the transition intensity rate matrix $\mathbf{Q}$;\\
  6. Reiterate 3-5 until the final iteration reached. Backtrack the serial numbers of changepoint locations.\\

  We may call the tuning parameter $k$ in Algorithm 3 as a resolution parameter. By choosing large $k$, a relatively large number of uniform times will be generated and the FFBS algorithm is carried out in a fine time scale, leading to accurate estimation of changepoints. However, with large number of uniformization events, both the memory costs and the computational costs will inflate quadratically. Nevertheless, choosing a small $\lambda(t)$ in the uniformization scheme will lower the memory costs and the computational costs, but the FFBS algorithm will be implemented in a coarser time scale, reducing the accuracies of the estimation of changepoints. The tuning parameter needs to be carefully selected for a tradeoff between the accuracies and efficiencies.

   Even after pruning the redundant changepoints and squeezing the controlling Poisson intensity rate used in uniformization scheme,  it is still possible that a large number of uniform times are occasionally generated by a "moderate" size of $\lambda(t)$, causing sharply inflated memory costs and computational costs. In this case, we consider a truncated version of uniformization scheme by setting an upper bound for the number of Poisson events generated from the uniformization procedure. FFBS algorithm is carried out only for those uniformizations with the number of Poisson events below this bound. The truncation error is given by

   \begin{equation}
     \|\bar{\mathbb{P}}_{s,t}^L-\mathbb{P}_{s,t}\| \le \sum\limits_{k=L+1}^{\infty} \frac{\lambda^k(t-s)^k}{k!}e^{-\lambda(t-s)} \le \frac{\lambda^L(t-s)^L}{L!},
   \end{equation}
   where $\|.\|$ denotes the standard supremum norm and $\bar{\mathbb{P}}_{s,t}^L$ is the truncated version of (3) at level $k=L$, see Van Dijk (2018).

  The above scheme is applicable for modelling any multiple changepoint models once the marginal likelihood of it can be exactly evaluated, which is evident when the conjugate priors are elicited for $\theta$, e.g. for multiple changepoint models from exponential families, or the marginal likelihood of the model can be numerically computed.

\section{MAP Estimation via the Continuous-time Viterbi Algorithms}

  The maximum a posteriori estimates of the number of changepoints and their locations can be obtained from continuous-time Viterbi algorithm. Define
  \begin{equation}
  M_i(t_k)=\max\limits_{\mathbf{X}(t_1):\mathbf{X}(t_{k-1})}\log L(\mathbf{X}(t_1):\mathbf{X}(t_{k-1}),\, \mathbf{X}(t_k)=i,\, y_{1:k}), k=1,\cdots,n,
  \end{equation}
  where $\mathbf{X}(t_1):\mathbf{X}(t_{k-1})$ denotes $(\mathbf{X}(t_1), \cdots, \mathbf{X}(t_{k-1}))$. The Viterbi recursion shows that
  \begin{eqnarray}
  M_j(t_{k+1})=\max\limits_i\{M_i(t_k)+\log p_{ij}(t_{k+1}-t_k)\} + \log p(y_{k+1},\theta_j),
  \end{eqnarray}
  where $p_{ij}(t_{k+1}-t_k)$ is the likelihood of the latent Markov chain on $(t_k,t_{k+1}]$ with $\mathbf{X}(t_k)=i$ and $\mathbf{X}(t_{k+1})=j$. Now, $p_{ij}(t_{k+1}-t_k)$ needs to be maximized over the sample path space. Note that $\mathbf{X}(t)$ has at most one jump between two consecutive observations at $t_k$ and $t_{k+1}$. When $j=i$, $\log p\big(\mathbf{X}(t_{k+1})=j\big|\mathbf{X}(t_k)=i\big)$ is a constant. Only $\log p\big(\mathbf{X}(t_{k+1})=i+1\big|\mathbf{X}(t_k)=i\big)$ needs to be maximized over the path space. Assume that the sample path of $\mathbf{X}(t)$ over $(t_k, t_{k+1}]$ is given by $\mathbf{X}(t_k,u)=i, \mathbf{X}[u, t_{k+1}]=i+1.$ The probability density is given by $q_ie^{-q_i(u-t_k)}e^{-q_{i+1}(t_{k+1}-u)},$ which is maximized by setting either $u-t_k$ or $t_{k+1}-u$ to zero. So the maximum of it is given by
  $q_i\max\{e^{-q_i(t_{k+1}-t_k)},e^{-q_{i+1}(t_{k+1}-t_k)}\} =q_ie^{-\min\{q_i,\, q_{i+1}\}(t_{k+1}-t_k)}$.
  Hence, the latent Markov chain has state transitions only at event times  of $\mathbf{Y}(t)$.

  \newtheorem{theorem}{Proposition}
  \begin{theorem}
   The optimal posterior path of the latent Markov chain $\mathbf{X}(t)$ given $\mathbf{Y}(t), 0\le t \le T$ has jumps only at event times.
  \end{theorem}

  According to the above proposition, the MAP estimates of the set of changepoints are selected exactly from the auxiliary event times, which can be implemented by a discrete-time version of Viterbi algorithm (Viterbi, 1967).  The dynamic programming algorithm performs on an event-by-event scale and the memory cost in this case is quadratic in $n$, which is prohibitive for long sequence of observations. Instead, we may first obtain the MAP estimates of the number of changepoints directly from the marginal posterior of the number of changepoints. Then, conditioned on the number of changepoints, the MAP estimates of the locations of changepoints can be obtained from the continuous-time version of Viterbi algorithm for HMMs on a finite state space, see also (Bebbington, 2007).

   \textbf{Algorithm 4} (Continuous-time Vierbi Algorithm)\\
  1. Initialize $f_1(j)=\pi' e^{\mathbf{Q}\Delta t_1}\Upsilon(y_1)[,j]$ and set $\phi_1(j)=0$ for all $j$, where $\Upsilon(y)= \mbox{diag}(p_1(y),\cdots,p_{m+1}(y))$ and $\pi'=(1,0,\cdots,0)$.\\
  2. For $k=2,\cdots,n$ and all $j$, recursively compute
  \begin{equation*}
   f_k(j)=\max\limits_i\{f_{k-1}(i)e^{\mathbf{Q}\Delta t_k}\Upsilon(y_k)[i,j]\}
  \end{equation*}
  and
  \begin{equation*}
  \phi_k(j)=argmax_i\{f_{k-1}(i)e^{\mathbf{Q}\Delta t_k}\Upsilon(y_k)[i,j]\}.
  \end{equation*}\\
  3. Let $x_n=argmax_jf_n(j)$ and backtrack the state sequence as follows:\\
     For $k=n-1, \cdots, 1, x_k=\phi_{k+1}(x_{k+1}).$\\

    In the above algorithm, $f_k(j)$ propagates to zero or infinity exponentially fast, which will cause underflow or overflow problem. Proper scaling procedure by taking logarithm of it or other approaches are required in numerical computations.

\section{Simulation Studies}

   We perform a simulation study to assess the performance of the method for a sequence of 1200 Normally distributed observations with 10 changepoints. The sequence of observations contain a variety of jump sizes in mean and a variety of segment lengths. The top of Figure 1 demonstrates the true signal of the noisy observations and the exact changepoint locations. For simplicity, we assume the changepoints appear only in the mean. The marginal likelihood of the Normal model $\mathcal{N}(y|\mu, \sigma^2)$ for $y_s, \cdots, y_t$ is given by
  \begin{eqnarray}
  & & \int \prod_{i=s}^{t} \mathcal{N}\left(y_{i} | \mu, \sigma^{2}\right) \mathcal{N}\left(\mu | m, \tau^{2}\right) d \mu \\\nonumber &=&\frac{\sigma}{(\sqrt{2 \pi} \sigma)^{r} \sqrt{r \tau^{2}+\sigma^{2}}} \exp \left(-\frac{\sum\limits_{i=s}^t y_{i}^{2}}{2 \sigma^{2}}-\frac{m^{2}}{2 \tau^{2}}\right) \exp \left(\frac{\frac{\tau^{2} r^{2} \overline{y}_{s:t}^{2}}{\sigma^{2}}+2 r \overline{y}_{s:t} m}{2\left(r \tau^{2}+\sigma^{2}\right)}\right),
  \end{eqnarray}
  where $\mathcal{N}(\mu|m, \tau^2)$ is the conjugate prior of $\mu$. In (14), $\sigma^2, \tau^2$ and $m$ are known constants, $r=t-s+1$ and $\overline{y}_{s:t}$ is the segment mean. Assume a prior $\mathcal{N}\left(\mu | 1.5, 1\right)$ for $\mu$ and let $\sigma^2=1$. The segment means are $c(0,3,0,2,0,-2,0,3,0,3,0)$. The true positions of changepoints are 101, 161, 261, 361, 481, 601, 701, 801, 901 and 1001. We attach auxiliary event times for discrete observations on a time interval $[0,T]$ to facilitate the use of uniformization scheme. Without loss of generality, the observations are arranged just regularly on $[0,T]$. In this simulation, Gibbs sampler iterates 6000 times, with the last 3000 samples collected as independent draws.

  In the uniformization scheme, the Poisson intensity rate is set to be k times of the changepoint recurrence rate $q_i$ in $\mathbf{Q}$. To control the memory cost and the computational cost, we set an upper bound 250 for the number of Poisson event numbers generated from the uniformization scheme. The truncation error can be evaluated by (11). This upper bound is rarely reached for k below 15. The truncation error can be neglected completely. In this numerical example, we demonstrate the scaling effects on the accuracies and efficiencies of the estimation. Figure (2) indicates the posterior of the number of changepoints under three different scalings $k= 5, 11$ and 15. It is observed that the posterior of the number of changepoints is more concentrated for large $k$, but more diffused for small $k$ instead. In this case, it is prone to underestimate the number of changepoints for small scaling $k$. Further evidence of the scaling effect on the accuracies of the estimation is demonstrated in Figure (3). Figure (3) indicates the posterior of the locations of changepoints under three scalings. From Figure 3, it is observed that the posterior of the locations of changepoints is rather dispersed upon $k=5$. In this case, the upper (lower) limits of $95\%$ HPD intervals of changepoint locations are partially superimposed with adjacent ones and the bias of the estimates of changepoint locations from posterior means is large. After increasing $k$, the $95\%$ HPD interval of changepoint locations are narrowed and the bias of the estimates of changepoint locations is decreased. The estimation is reasonably accurate for $k=15$. Further increasing $k$ is unnecessary as the memory cost and the computational cost would inflate quadratically with respect to the number of uniform events and the accuracies of the estimation will not be improved significantly.

  In simulations, we find that there still exist remaining "knots" in the simulated changepoints even after pruning. Generally, the posterior of the number of changepoints summarized from this scheme tends to overestimate the number of changepoints. A better summary of the posterior of the number of changepoints is obviously indicated in the posterior of the locations of changepoints, since these "knots" will be merged into the posterior histogram (density) estimation by choosing an appropriate bandwidth. In this case, the number of changepoints can be summarized by counting the number of bumps in the histogram (density), see Figure 3 and Figure 2. In the middle of Figure 7, it is clear that there exists 10 bumps in the posterior. Now, the locations of changepoints can be summarized either by the posterior means or the MAP estimates via continuous-time Viterbi algorithm. The MAP estimate of changepoint locations via CT-Viterbi algorithm in this case is 100, 160, 258, 360, 478, 603, 700, 802, 900 and 1000, which is very accurate. However, the posterior mean estimates of changepoint locations is given by 114, 183, 278, 377, 497, 620, 718, 818, 916 and 1012, which is only reasonably close to the true values, due to the existence of knots and other factors. The posterior mean estimate of $\mu$ is given by 0.42, 2.36, 0.33, 1.54, -0.20, -1.67, 0.56, 2.45, 0.47, 2.58 and 0.21, which is also close to the true values. Figure 4 demonstrates the $95\%$ HPD credible intervals for some $\mu_i$s with a kernel density estimation.

  The second simulation is to display sparse changepoint detection for an exponentially distributed sequence in length $n=10000$ with just 3 changepoints located at 3000, 5000 and 7000. We suggest the uniformization scheme is potentially suitable for changepoint detection for a long sequence of observations when the changepoints are only sparsely distributed. In this numerical example, it will be extremely expensive to locate the changepoints in an event-by-event scale, as both the memory cost and the computational cost are quadratic to the number of events. To mitigate the sharply inflated memory costs and computational costs, one may consider to group the observations into a few hundred of blocks and detect the locations of changepoints in a block-by-block scale. However, the blocking error can be unignorable if the block size is large, causing a poor estimate of the locations of changepoints. On the contrary, choosing a small block size to control the blocking error will inflate the memory cost and the computational cost quickly. By randomized blocking via an auxiliary uniformization, there is no blocking error to be quantified and we still can locate changepoints accurately at much lower memory costs and computational costs.

  The marginal likelihood of the exponential model $\exp(\lambda)$ for $y_s, \cdots, y_t$ with a conjugate prior $\Gamma(\alpha,\beta)$ is given by
  \begin{eqnarray}
  & & \int \prod_{i=s+1}^{t} Exp\left(y_{i} | \lambda\right) \Gamma\left(\lambda | \alpha, \beta\right) d \lambda \\\nonumber
  &=&\frac{\beta^\alpha}{\Gamma(\alpha)}\frac{\Gamma(t-s+\alpha)}{\left(\sum\limits_{i=s+1}^{t}y_i+\beta\right)^{t-s+\alpha}}.
  \end{eqnarray}
  In this simulation, Gibbs sampler iterates 1500 times, with the last 500 samples collected as independent draws. In the uniformization scheme, the Poisson intensity rate is set to be 10 times of the changepoint recurrence rate in $\mathbf{Q}$. To alleviate the memory cost and the computational cost, we set an upper bound 250 for the number of Poisson events generated from the uniformization scheme. In simulations, this bound is rarely reached and the truncation error can be neglected. The top of Figure 5 demonstrates the centralized and normalized cumulative sums for the simulated observations, which is given by $\frac{\sum_{i=1}^{j}y_i}{\sum_{i=1}^{n}y_i}-\frac{j}{n}$. For a stationary process, the statistic should be close to the line segment: $y=0, 0\le x \le 1$ and behave like a Brownian bridge on $[0,1]$. We stretch the x-axis from $[0,1]$ to $[0,n]$. From the top of Figure 5, it is visible that there exists roughly 3 changepoints in the observations. This kind of exploratory data analysis is applicable for choosing appropriate initialization parameters in Algorithm 3. The middle of Figure 5 shows the histogram of the locations of changepoints. With 3 "bumps" indicated in the figure, it can be deduced that there exists 3 changepoints, which is a more accurate indication of the number of changepoints than that indicated in the bottom of Figure 5. The bottom of Figure 5 is obviously less informative about the posterior of the number of changepoints due to existing "knots". It is noted that the algorithm occasionally break down at low resolution (small k), due to that all the filtering probabilities drop to zeros exponentially fast, causing underflow problems, which is difficult to deal with for the moment. In practice, it might be necessary to make some pilot runs for the selection of an appropriate resolution parameter.

\section{Real Numerical Examples}

     We perform a real data analysis for New Zealand deep earthquakes. The data set includes 674 events with magnitude greater than 5 in Richter scale, which is selected from New Zealand catalogue between 1945 and 2015 at depth greater than 45km within a polygon with vertices $(170^{\circ}E, 43^{\circ}S)$, $(175^{\circ}E, 36^{\circ}S)$, $(177^{\circ}E, 36^{\circ}S)$, $(180^{\circ}E, 37^{\circ}S)$,  $(180^{\circ}E, 38^{\circ}S)$, $(173^{\circ}E, 45^{\circ}S)$ and $(170^{\circ}E, 43^{\circ}S)$. The data set is freely available from GNS Science of New Zealand via Geonet (www.geonet.org.nz). The main occurrence pattern of New Zealand deep earthquakes is that it varies from time to time. It is active in one period, and relatively quiescent in another. Wether the time-varying behavior of the deep earthquakes is only a random fluctuation or it is associated with structural breaks of deep seismicity is important in seismic risk forecasting and hazard evaluation. We characterize the time-varying pattern of deep seismicity by a Poisson changepoint model. We count the total number of deep earthquakes in a quarter of one year (exactly three months) in a time span of 70 years and the counts of deep earthquakes are assumed from Poisson distributions. The marginal likelihood of the Poisson model $Pois(\lambda)$ for $y_{s+1}, \cdots, y_t$ with conjugate priors $\Gamma(\alpha,\beta)$ is given by
  \begin{eqnarray}
  & & \int \prod_{i=s+1}^{t} Pois\left(y_{i} | \lambda\right) \Gamma\left(\lambda | \alpha, \beta\right) d \lambda \\\nonumber
  &=&\frac{\beta^\alpha}{\Gamma(\alpha)}\frac{\Gamma\left(\sum\limits_{i=s+1}^{t}y_i+\alpha\right)}{\left(t-s+\beta\right)
  ^{\left(\sum\limits_{i=s+1}^{t}y_i+\alpha\right)}}\frac{1}{\prod\limits_{i=s+1}^{t}y_i!}.
  \end{eqnarray}

   Gibbs sampler iterates 5000 times, with the last 3000 draws treated as posterior samples. In the uniformization scheme, the Poisson intensity rate is set to be 15 times of the changepoint recurrence rate in $\mathbf{Q}$. We set an upper bound 250 for the number of Poisson event numbers generated from the uniformization scheme. This upper bound is never reached in the iterations and the truncation error can be neglected. The top of Figure 6 shows the quarterly counts of large deep earthquakes from 1945 to 2015. We look at whether the fluctuations of deep earthquake counts are random or related to deep seismicity change. The middle of Figure 6 indicates the posterior of changepoint locations. From the figure, it is observed that large uncertainties appear for the number of changepoints and their locations. Although the MAP estimates of the number of changepoints is 2, we deduce there exists at least two or three changepoints in the deep seismicity rates. The posterior means of changepoint locations and their $95\%$ HPD intervals for three changepoints are indicated in the figure. The two most likely changepoint locations obtained via continuous-time Viterbi algorithm are 1986 and 2007. It seems that the deep seismicity is quiescent before 1986 and after 2007, but relatively active between 1986 and 2007.

   The second real data analysis is performed for the well-log data. The data set consists of 4050 observations from nuclear magnetic response from a drill head drilling through the rock strata. The signal is piecewise constant with white noise and some outliers. Each segment of the signal is related to a rock type. In this analysis, the outliers are removed manually and the method is applied to the remaining 3957 measurements, see the top of Figure 7 for the rescaled data with outliers manually removed. The reader is referred to Fearnhead and Rigaill (2019) for the changepoint detection in the presence of outliers. We assume a Normal model $N(y|\mu_t, \sigma^2)$ for this sequence of observations , where $\mu_t$ is the mean of the signal and the variance $\sigma^2$ is assumed known. For simplicity, the observations are rescaled to follow nearly a standard Normal distribution. Let $y'=(y-115000)/10000$. Assume $\sigma^2=1$ and set a conjugate prior $N(0,1)$ for $\mu_t$. The marginal likelihood for a segment is indicated in (14).

   In this analysis, the Gibbs sampler in Algorithm 3 run for 1000 burn in followed by 1000 iterations, which are collected as posterior samples. In the uniformization scheme, the Poisson intensity rate is $12$ times of the changepoint recurrence rate as given in $\mathbf{Q}$. To limit the memory cost and the computational cost, the number of Poisson events generated from uniformization scheme is upper bounded by 250, which is rarely reached and the truncation error can be neglected. The bottom of Figure 7 indicates the posterior of the number of changepoints. The MAP estimate of the number of changepoints is 10. The posterior of the locations of changepoints are demonstrated in the middle of Figure 7, with the posterior mean of changepoint locations and their $95\%$ HPD intervals indicated in the Figure. The MAP estimates of changepoint locations are given in the top of Figure 7. Visibly,  the MAP estimates of changepoint locations are better than posterior means.

\begin{small}

\end{small}

Figure 1: The figure demonstrates the true signal of the noisy observations by red solid lines. The exact changepoint locations are indicated by vertical dash lines.\\

Figure 2: The three figures display the posteriors of the number of changepoints. The posteriors of the number of changepoints are indicated for different values of the scaling parameter k at three levels, e.g. $k=5, 11$ and 15 in the top, the middle and the bottom of the figure respectively.\\

Figure 3: The three figures display the locations of changepoints for different values of the scaling parameter k at three levels, e.g. $k=5, 11$ and 15 in the top, the middle and the bottom of the figure respectively. In each figure, the posterior means and the 95\% HPD credible intervals are demonstrated by red dash lines and red sold lines respectively, with the exact positions of 3 changepoints indicated by blue solid lines.\\

Figure 4: Kernel density estimates for part of the model parameters. Line segments in bold beneath each kernel density estimate indicate the $95\%$ highest posterior density for the model parameters. The left top, right top, left bottom and right bottom of the figure show the posterior summary of $\mu_1, \mu_2, \mu_3 \text{and} \mu_4$ respectively. The number of posterior samples and the Bandwidth used in the kernel density estimation are given.\\

Figure 5: The top of the figure displays the centralized and normalized cumulative sum of the observations. The middle of the figure shows the histogram of changepoint locations from the posterior samples. The bottom of the figure shows the posterior of the number of changepoints.\\

Figure 6: The top of the figure displays the quarterly counts (about 90 days) of deep earthquakes in New Zealand. The middle of the figure shows the histogram of changepoint locations from the posterior samples, with the posterior means and the 95\% HPD credible intervals demonstrated by red solid lines. The bottom of the figure shows the posterior for the number of changepoints.\\

Figure 7: The top of the figure displays the Well-log data with outliers removed manually. the solid lines indicating the true signals and the vertical dash lines indicating the MAP estimates of changepoint locations by Viterbi algorithms. The middle of the figure shows the histogram of changepoint locations from the posterior samples, with the posterior means and the 95\% HPD credible intervals demonstrated by red solid lines. The bottom of the figure shows the posterior of the number of changepoints.\\

  \begin{figure}
  \begin{center}
  \rotatebox{0}{\scalebox{1}[1]{\includegraphics{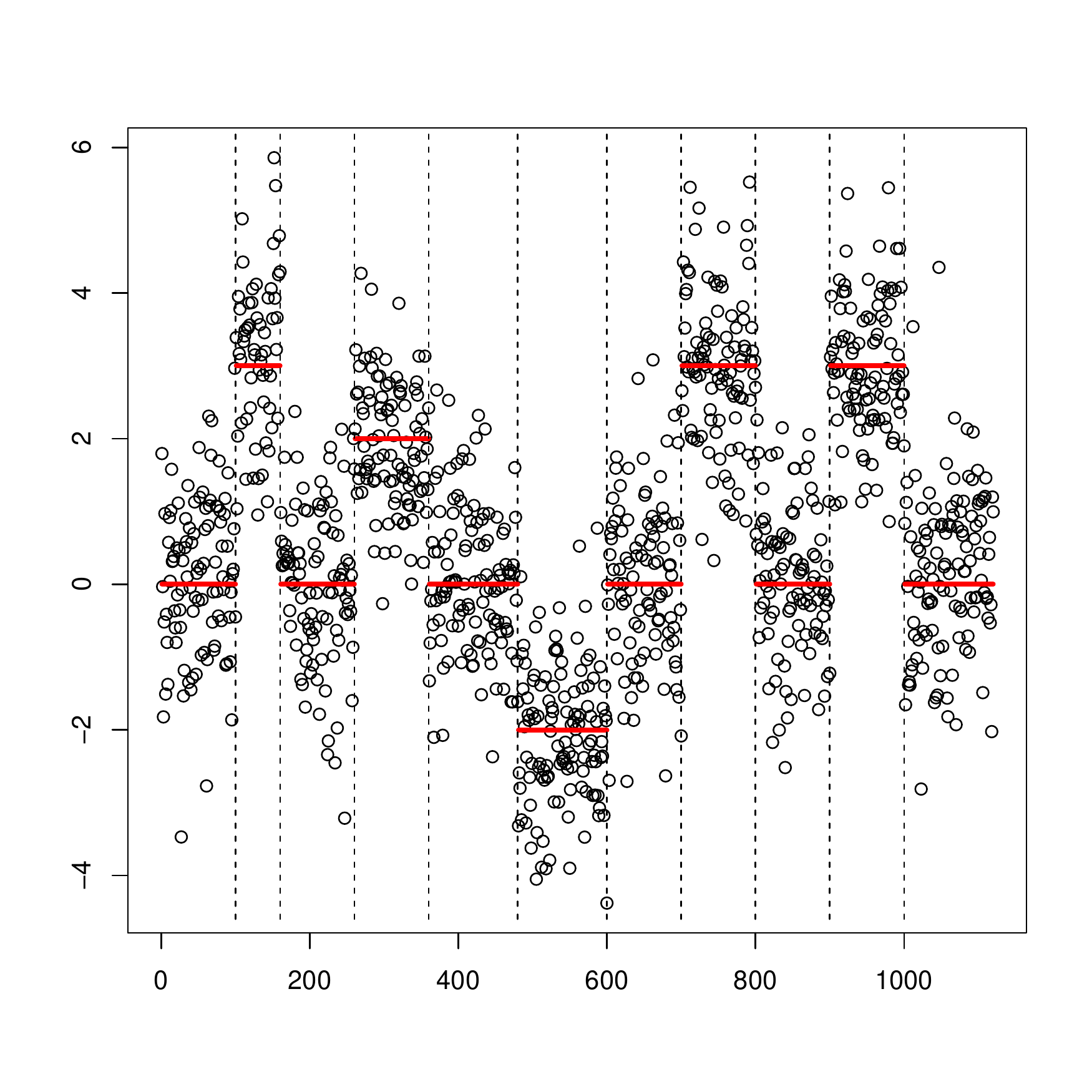}}}
  \end{center}
  \caption{}
  \end{figure}

  \begin{figure}
  \begin{center}
  \rotatebox{0}{\scalebox{1}[1]{\includegraphics{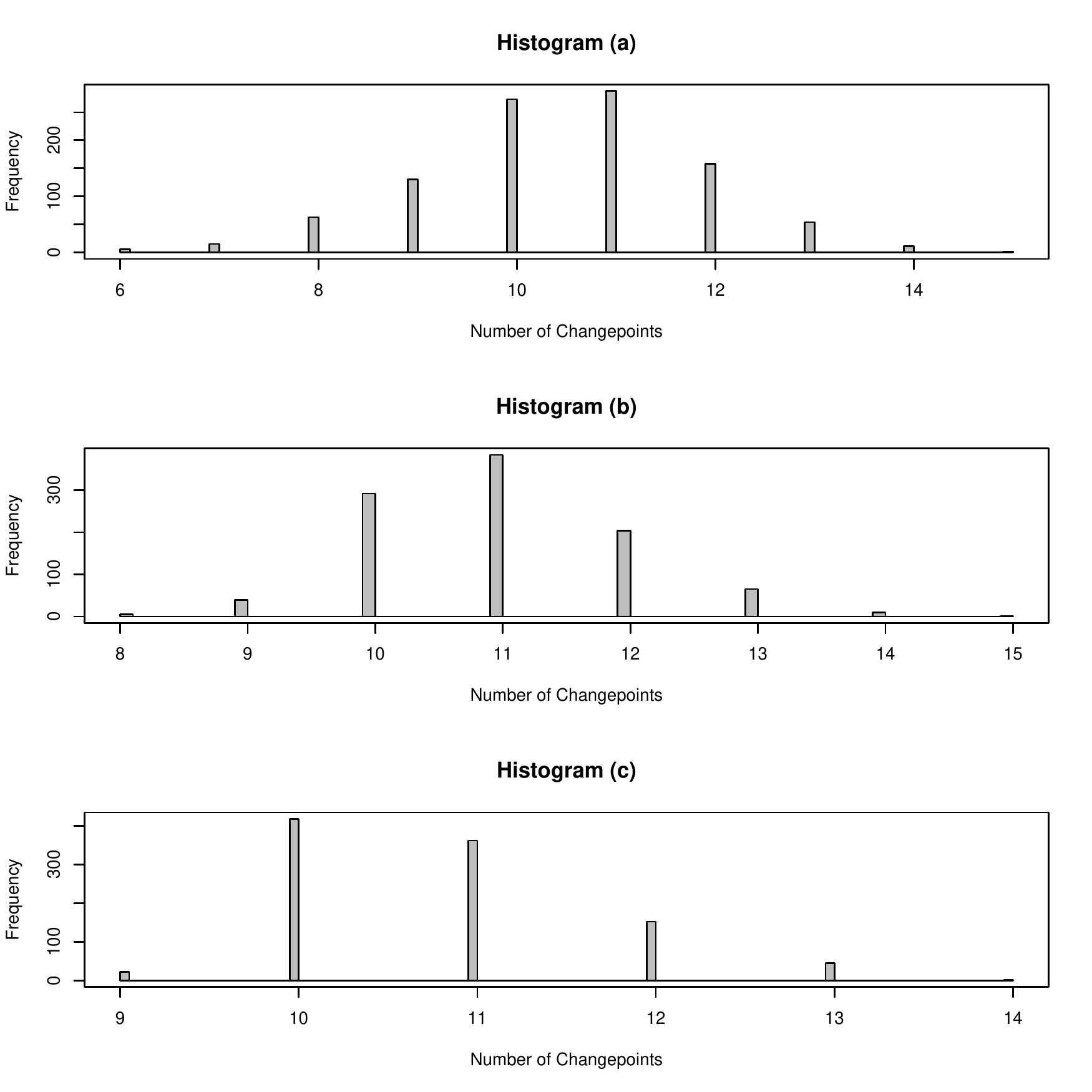}}}
  \end{center}
  \caption{}
  \end{figure}

  \begin{figure}
  \begin{center}
  \rotatebox{0}{\scalebox{1}[1]{\includegraphics{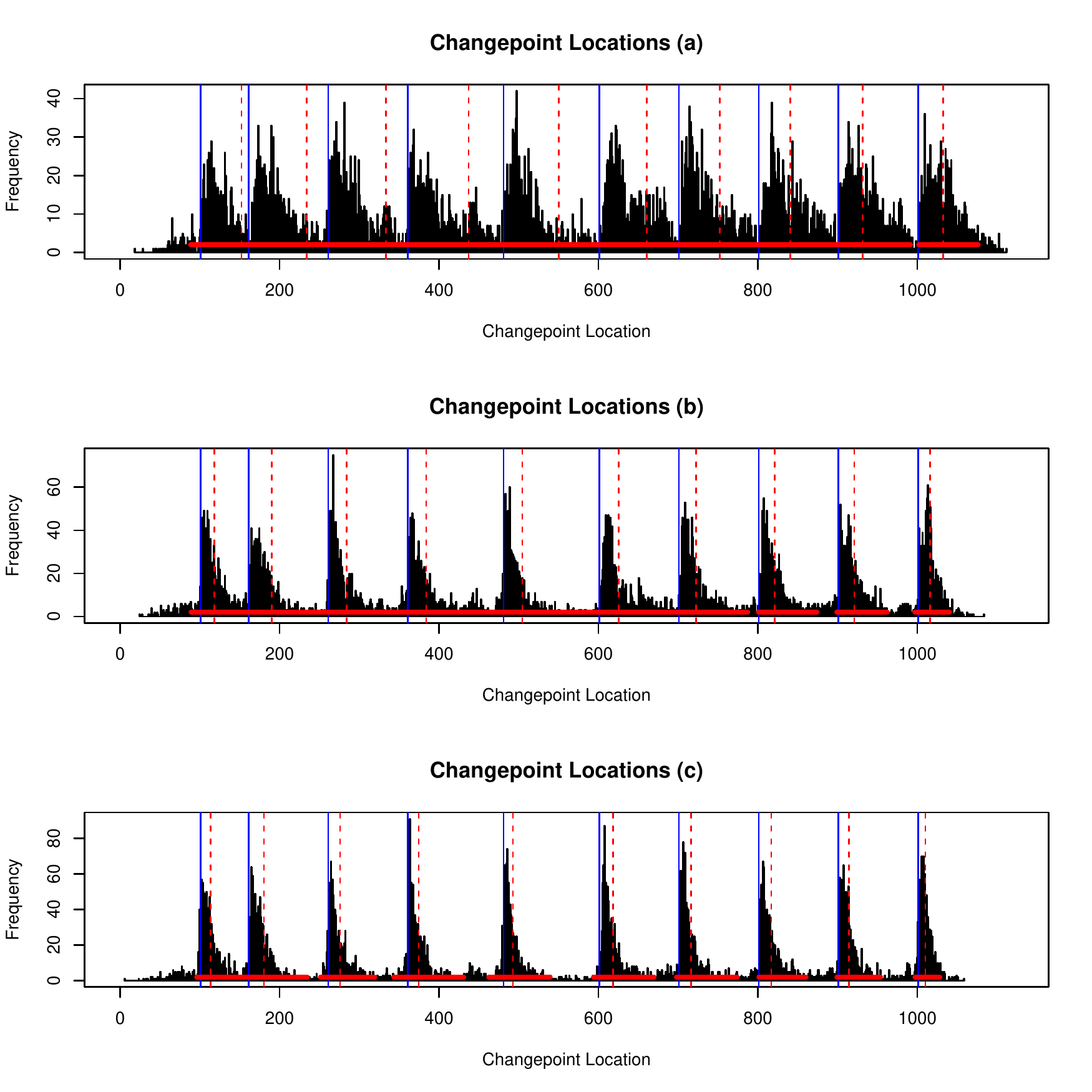}}}
  \end{center}
  \caption{}
  \end{figure}

  \begin{figure}
  \begin{center}
  \rotatebox{0}{\scalebox{1}[1]{\includegraphics{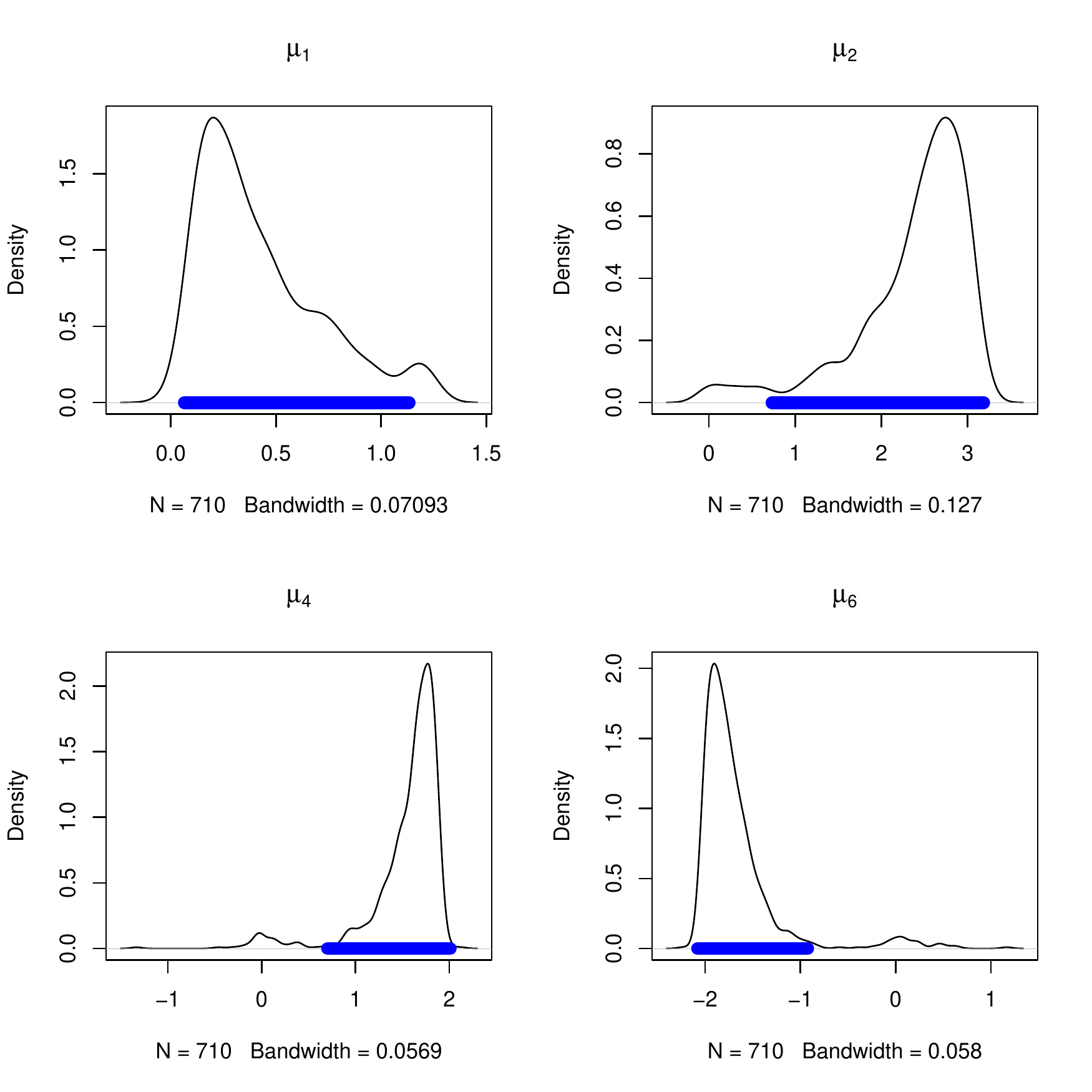}}}
  \end{center}
  \caption{}
  \end{figure}

  \begin{figure}
  \begin{center}
  \rotatebox{0}{\scalebox{1}[1]{\includegraphics{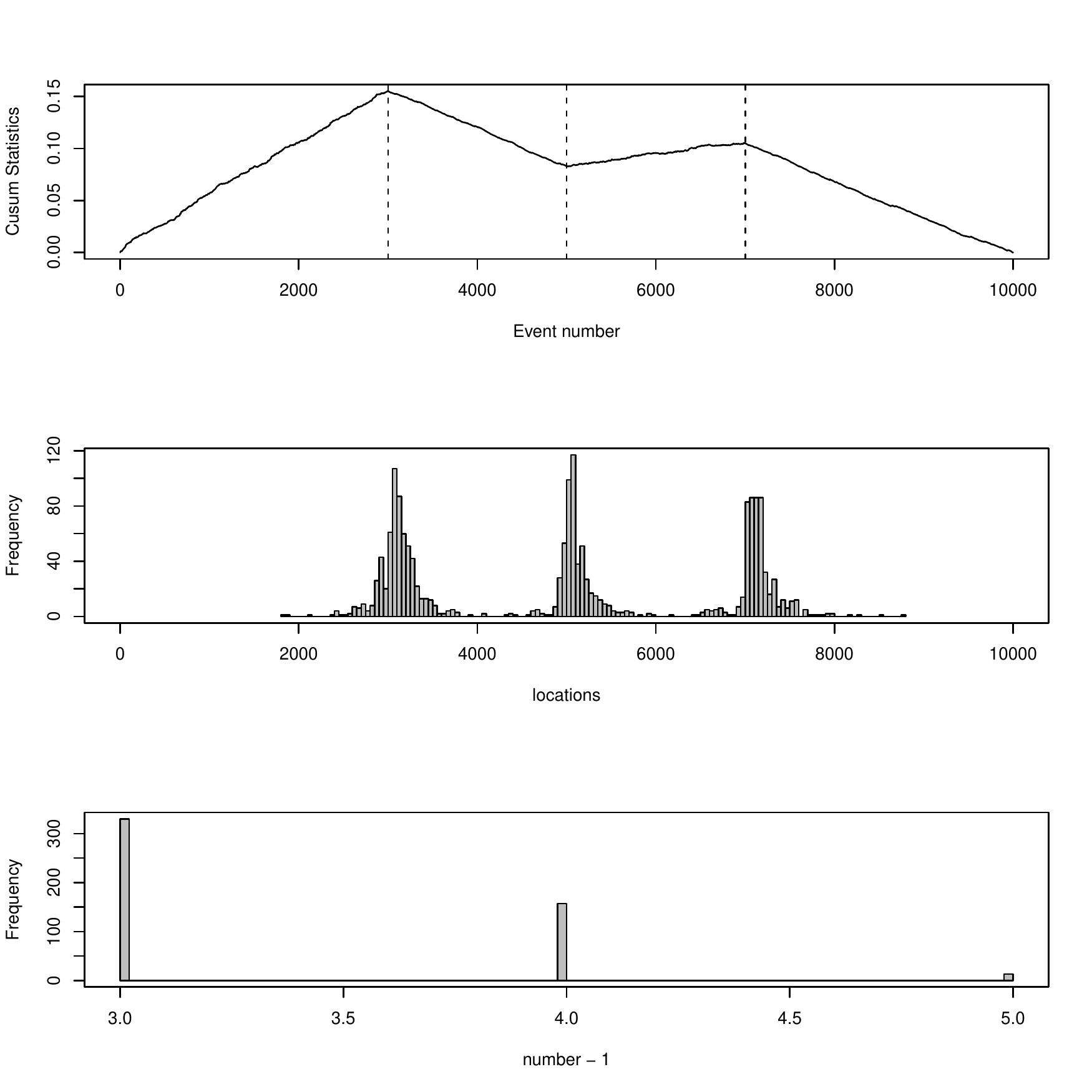}}}
  \end{center}
  \caption{}
  \end{figure}

  \begin{figure}
  \begin{center}
  \rotatebox{0}{\scalebox{1}[1]{\includegraphics{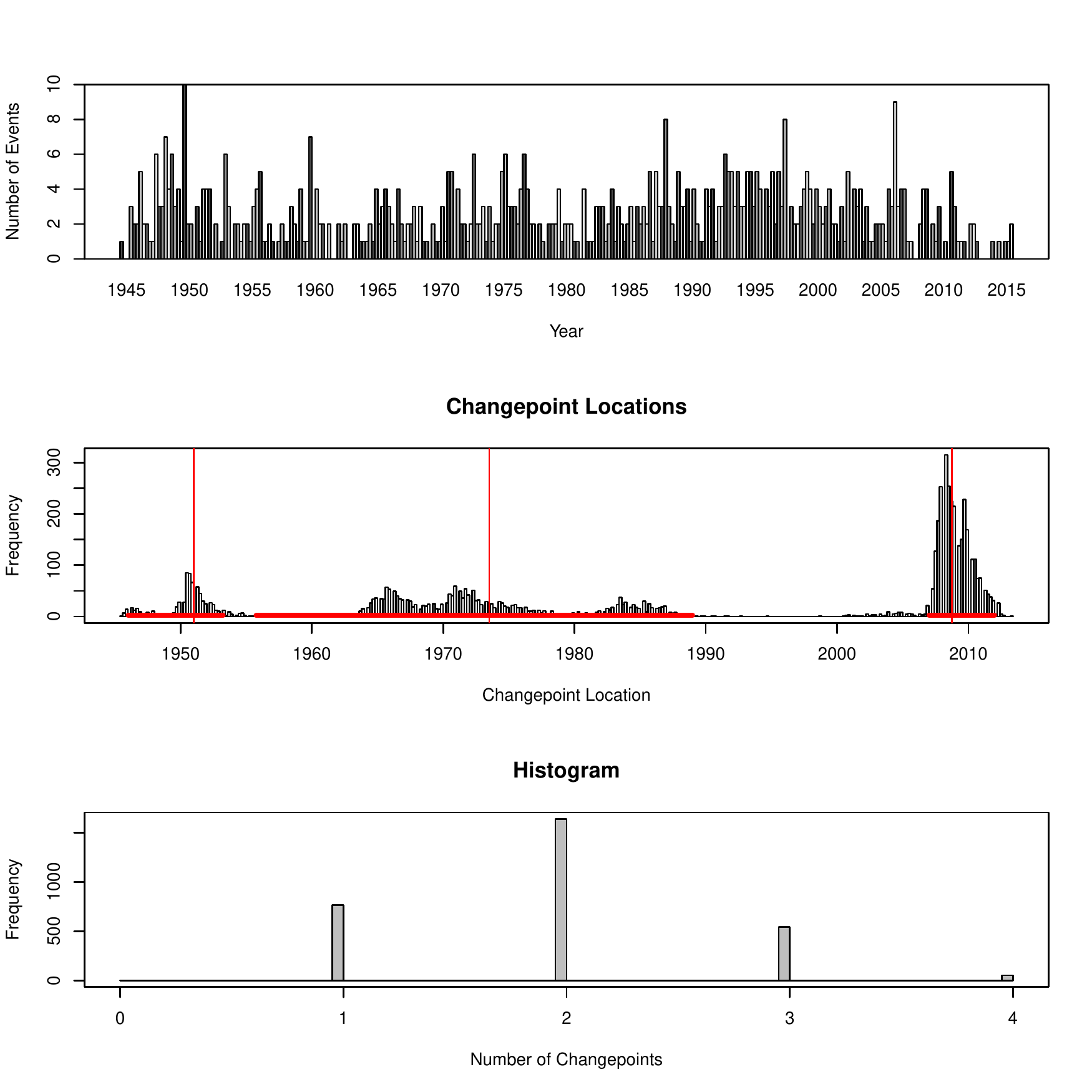}}}
  \end{center}
  \caption{}
  \end{figure}

  \begin{figure}
  \begin{center}
  \rotatebox{0}{\scalebox{1}[1]{\includegraphics{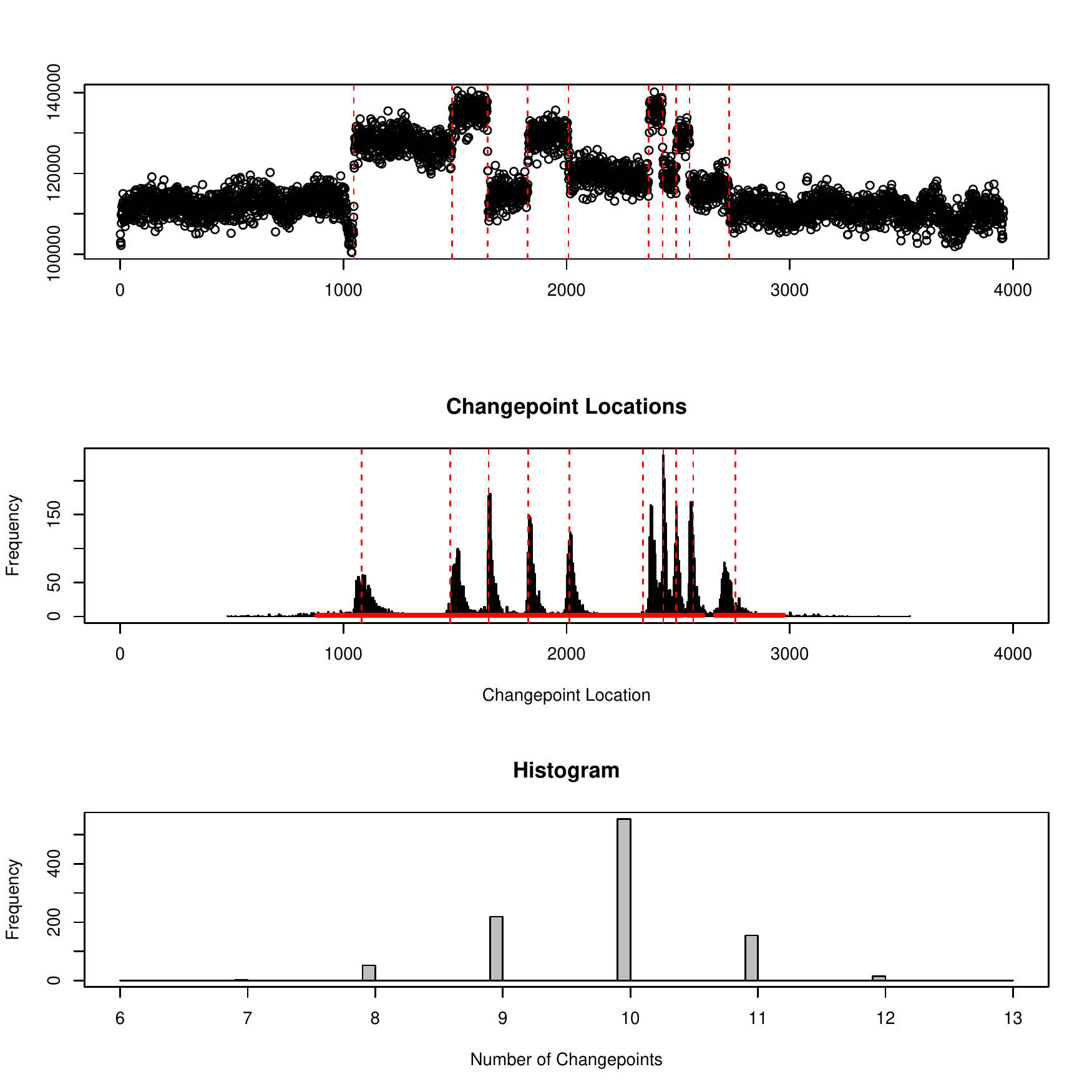}}}
  \end{center}
  \caption{}
  \end{figure}

\end{document}